\documentclass[a4paper,fleqn,usenatbib]{mnras}

\usepackage[T1]{fontenc}
\usepackage{ae,aecompl}

\usepackage{graphicx}	
\usepackage{amsmath}	
\usepackage{amssymb}	
\usepackage{xspace}	
\usepackage{comment}
\usepackage{booktabs}

\newcommand{\msun}{{\rm M}_{\odot}}
\newcommand{\lsun}{{\rm L}_{\odot}}
\newcommand{\rsun}{{\rm R}_{\odot}}

\newcommand{\bonnsai}{\mbox{\textsc{Bonnsai}}\xspace}
\newcommand{\binamics}{\mbox{BinaMIcS}\xspace}
\newcommand*{\eg}{e.g.\@\xspace}
\newcommand*{\ie}{i.e.\@\xspace}
\newcommand{\hl}[1]{#1}

\title[Origin of magnetic fields in massive stars]{Rejuvenation of stellar mergers and the origin of magnetic fields in massive stars}

\author[Fabian~R.N.~Schneider~et~al.]{
F.R.N. Schneider$^{1}$\thanks{E-mail: fabian.schneider@physics.ox.ac.uk},
Ph. Podsiadlowski$^{1}$,
N. Langer$^{2}$,
N. Castro$^{2}$,
and L. Fossati$^{3}$
\\
$^{1}$Department of Physics, University of Oxford, Denys Wilkinson Building, Keble Road, Oxford OX1~3RH, United Kingdom\\
$^{2}$Argelander-Institut f{\"u}r Astronomie der Universit{\"a}t Bonn, Auf dem H{\"u}gel~71, 53121~Bonn, Germany\\
$^{3}$Space Research Institute, Austrian Academy of Sciences, Schmiedlstrasse 6, 8042 Graz, Austria
}

\date{Accepted XXX. Received YYY; in original form ZZZ}

\pubyear{2015}

\begin{document}
\label{firstpage}
\pagerange{\pageref{firstpage}--\pageref{lastpage}}
\maketitle

\begin{abstract}
\hl{Approximately 10\% of massive OBA main-sequence (MS) and pre-MS stars harbour strong, large-scale magnetic fields. At the same time there is a dearth of magnetic stars in close binaries. A process generating strong magnetic fields only in some stars must be responsible with the merging of pre-MS and MS stars being suggested as one such channel. Stars emerging from the coalescence of two MS stars are rejuvenated, appearing younger than they are. They can therefore be identified by comparison with reference clocks. Here we predict the rejuvenation of MS merger products over a wide range of masses and binary configurations calibrated to smoothed-particle-hydrodynamical merger models. We find that the rejuvenation is of the order of the nuclear timescale and is strongest in the lowest-mass mergers and the most evolved binary progenitors with the largest mass ratios. These predictions allow us to put constraints on the binary progenitors of merger products. We show that the magnetic stars HR~2949 and $\tau$~Sco are younger than the potential binary companion HR~2948 and the Upper-Sco association, respectively, making them promising merger candidates. We find that the age discrepancies and the potential binary progenitors of both are consistent with them being rejuvenated merger products, implying that their magnetic fields may originate from this channel. Searching for age discrepancies in magnetic stars is therefore a powerful way to explore which fraction of magnetic stars may have obtained their strong magnetic field in MS mergers and to improve our understanding of magnetism in massive stars and their remnants.}
\end{abstract}




\section{Introduction}\label{sec:introduction}

About 7--10\% of main-sequence (MS) OBA stars host strong, large-scale magnetic fields (\eg \citealt{2009ARA&A..47..333D}; \hl{\citealt*{2015SSRv..tmp...12F}}; \citealt{2015A&A...582A..45F}) and a similar B-field incidence is found in pre-MS Herbig Ae/Be stars \citep{2007MNRAS.376.1145W,2013MNRAS.429.1001A}. Among these, Babcock's star (HD~215441), a chemically peculiar A0 star with a surface polar magnetic field strength of $34\,\mathrm{kG}$, is the current record holder. The strong field pre-MS stars show similar magnetic fluxes to those found in magnetic\footnote{In the following, we use the term ``magnetic stars'' for stars with strong, large-scale surface magnetic fields. Sub-gauss surface magnetic fields have now been observed as well in some stars \citep[\eg][]{2009A&A...500L..41L,2011A&A...532L..13P} and may be ubiquitous.} OBA stars, indicating a common origin. Interestingly there is a dearth of magnetic stars in close binaries, first noted by \citet{2002A&A...394..151C} in Ap stars and now confirmed by the \binamics project in more massive OB stars: the magnetic incidence in close, massive binaries is no more than about 2\% \citep{2013EAS....64...75N,2015IAUS..307..330A,2015arXiv150200226N}. As of today there is only one binary with a magnetic field detection in both components, the eccentric ($e=0.27$), $4.56\,\mathrm{d}$ B-star binary $\epsilon$~Lupi \citep{2015MNRAS.454L...1S}.

The origin of these strong magnetic fields is still unknown but from the rather low magnetic-star incidence of the order of 10\% it is clear that something special must have happened to these stars. In the literature, three formation channels are typically discussed: (i) strong B-fields inherited from the star forming cloud \citep[\eg][]{2001ASPC..248....3M,2001ASPC..248..305M}, (ii) a dynamo generating strong magnetic fields and (iii) mergers of pre-MS and/or MS stars creating B fields because of strong shear (\citealt{2009MNRAS.400L..71F,2012ARA&A..50..107L,2014IAUS..302....1L}; \hl{\citealt*{2014MNRAS.437..675W}}). Scenarios (i) and (ii) have difficulties explaining why only some stars show strong, large-scale surface magnetic fields and not all of them given that they formed from the same molecular cloud and are governed by the same physical processes, respectively. 

Neutron stars (NSs) and white dwarfs (WDs) also come in a highly and a less magnetic group. There is the class of highly magnetised neutron stars, so-called magnetars, with inferred surface magnetic fields of about $10^{13}\text{--}10^{15}\,\mathrm{G}$ \citep{2014ApJS..212....6O} and the class of strong-field WDs, among them the Polars, with surface magnetic fields in excess of $10^6\,\mathrm{G}$. \hl{\citet*{2003AJ....125..348L}} find that at least about $10\%$ of WDs with B-field strengths \hl{greater than about} $2\,\mathrm{MG}$ are highly magnetic. Because of the similar fractions of magnetic WDs and AB stars, \citet{1992A&ARv...4...35L} propose that the magnetic AB stars may be the progenitors of magnetic WDs. 

No highly magnetic WD has been found in a detached binary. Strong magnetic fields have only been found in WDs in cataclysmic variables \hl{(Roche-lobe overflowing binaries)} and in some apparently single WDs \citep{2008MNRAS.387..897T}. Based on that, \citet{2008MNRAS.387..897T} suggest that the strong magnetic fields in WDs are the result of a dynamo operating because of differential rotation and convection during common-envelope evolution that (nearly) lead to a merger. In this picture near mergers give rise to magnetic cataclysmic variables and complete mergers to highly magnetised single WDs. Following the work of \citet{2008MNRAS.387..897T}, \citet{2014MNRAS.437..675W} propose an $\Omega$ dynamo\footnote{\hl{Strictly speaking, this is not a pure $\Omega$ dynamo because \citet{2014MNRAS.437..675W} introduce an artificial mechanism that generates a poloidal field component from the decay of the toroidal component.}} that generates magnetic fields from differential rotation that could act during common-envelope evolution and in mergers. \citet{2014MNRAS.437..675W} further find that the magnetic flux per unit mass in highly magnetised WDs is similar to that of magnetic OBA stars and therefore suggest that the magnetic fields in strong field WDs and OBA stars may have both been generated by similar processes.

A scenario accounting for the observations mentioned above is that magnetic stars are produced by mergers of pre-MS and MS stars. Mergers can naturally explain why only a fraction of stars have strong, large-scale magnetic fields (\eg \citealt{2014ApJ...782....7D} predict that about 8\% of all massive MS stars are merger products), why the magnetic field characteristics of magnetic pre-MS stars, MS stars and strong field WDs are so similar and why there is a dearth of magnetic MS stars in close binaries. The merging can occur during the pre-MS by, \eg, tidal interaction with circumstellar material (\citealt{2010MNRAS.402.1758S}; \hl{\citealt*{2012A&A...543A.126K}}) as well as during the MS by, \eg, binary star evolution (\eg \hl{\citealt*{1992ApJ...391..246P}}; \citealt{2012ARA&A..50..107L}).

Main-sequence merger products are hard to distinguish from genuine single stars because their internal structure adjusts to the new mass such that it may be quite similar to that of single stars. However, mergers may show surface enhancements of hydrogen-burning products such as nitrogen and helium, they may be rapid rotators if not spun down efficiently, \eg by magnetic fields coupled to outflows, and there may be circumstellar material ejected prior to and/or during the merger \hl{process}. More importantly, MS merger products will be rejuvenated \citep[\eg][]{1983Ap&SS..96...37H,1992ApJ...391..246P,1995A&A...297..483B,2007MNRAS.376...61D}, \ie they will appear younger than other stars that formed at the same time such as other cluster members or binary companions \citep[\eg][]{1998A&A...334...21V,2014ApJ...780..117S}. Pre-MS mergers will not rejuvenate and will not show hydrogen-burning products on their surface. Apparent age discrepancies are therefore strong hints towards a MS merger origin of some magnetic stars and may further allow us to estimate the fraction of stars that obtained their strong magnetic field in MS mergers.

Age discrepancies have recently been reported in two magnetic stars. In the HR~2949 and~2948 visual binary, the more massive and magnetic star HR~2949 is found to be younger than the non-magnetic, potential binary companion HR~2948 \citep{2015MNRAS.449.3945S}. Also, the magnetic star $\tau$~Sco appears to be significantly younger than the Upper Scorpius association of which it is a proper motion member \citep{2014A&A...566A...7N}. Triggered by these findings, we investigate whether the merger hypothesis is able to explain the observed age spreads. To that end, we explore the rejuvenation of MS mergers (Sec.~\ref{sec:mergers}) and apply our findings to HR~2949 and $\tau$~Sco (Sec.~\ref{sec:merger-candidates}). We show that both stars may indeed be the products of MS mergers and put constraints on their progenitor systems. We discuss our results and further testable predictions of the merger hypothesis in Sec.~\ref{sec:discussion}, and summarize our conclusions in Sec.~\ref{sec:conclusions}.

\section{Rejuvenation of merger products}\label{sec:mergers}

We define the age of a star as the time that has passed since core hydrogen ignition (zero-age MS; ZAMS). A rejuvenated star is a star whose appearance suggests an apparent age that is younger than its real age\hl{. Rejuvenation can therefore} only be defined with respect to a comparison clock.

The merging of pre-MS stars leads to merger products that reach the ZAMS at different times compared to pre-MS genuine single stars of the same mass. The merger products may reach the ZAMS earlier because of the decreased contraction timescale due to the increased mass but it can also reach the ZAMS later because of the additional orbital energy that has to be radiated away before the core heats up enough to start hydrogen burning. The delay is, in any case, of the order of the thermal timescale and therefore practically undetectable.

In MS mergers the rejuvenation is of the order of the nuclear timescale and hence detectable. In the following, we predict the rejuvenation and hence apparent age differences of MS merger products to pave the way for the identification of MS mergers by apparent age discrepancies. To that end, we extend the approach of \citet{2008A&A...488.1017G} by also taking stellar wind mass loss into account which is important in massive stars. We assume that the average hydrogen mass fraction, 
\begin{equation}
\left< X \right> = \frac{1}{M} \int_0^M \, X(m) \,\mathrm{d}m = (1- Q_\mathrm{c} f) X_0,
\label{eq:avrg-x}
\end{equation}
decreases linearly with time from its initial value of $\left<X\right>_\mathrm{ZAMS}=X_0$ to its final value of $\left<X\right>_\mathrm{TAMS}$ at the terminal-age MS (TAMS). In Eqs.~(\ref{eq:avrg-x}), $M$ is the total mass and $m$ the mass coordinate, and we have defined the fractional MS age $f=t/\tau_\mathrm{MS}$ (with $t$ being the age and $\tau_\mathrm{MS}$ the MS lifetime of a star) and the effective core mass fraction,
\begin{equation}
Q_\mathrm{c} = \frac{\left<X\right>_\mathrm{ZAMS} - \left<X\right>_\mathrm{TAMS}}{\left<X\right>_\mathrm{ZAMS}} = \frac{X_0-\left<X\right>_\mathrm{TAMS}}{X_0}.
\label{eq:qc}
\end{equation}
In an ideal situation, \ie no wind mass loss, no receding convective cores during the course of the MS evolution, no additional mixing etc., the above defined effective core mass fraction gives $Q_\mathrm{c}=M_\mathrm{c}/M$ for massive stars with convective cores of mass $M_\mathrm{c}$. In non-ideal cases such as those encountered in this work, Eq.~\ref{eq:qc} gives the relative difference between the average initial and end-of-MS hydrogen mass fractions which can be viewed as an \emph{effective} core mass fraction (for simplicity we call $Q_\mathrm{c}$ the effective (convective) core mass fraction from here on). The big advantage of this definition is that it can also be applied to stars with radiative cores and it properly accounts for stellar wind mass loss, additional mixing, receding convective cores etc. 

We further assume that the stellar mass, $M$, decreases linearly with time because of winds,
\begin{equation}
M = (1 - Q_\mathrm{m} f) M_\mathrm{ini}.
\label{eq:present-day-mass}
\end{equation}
In the last equation, $M_\mathrm{ini}$ is the initial mass of a star and $Q_\mathrm{m}$ the fraction of mass lost on the MS,
\begin{equation}
Q_\mathrm{m} = \frac{M_\mathrm{ini} - M_\mathrm{TAMS}}{M_\mathrm{ini}}
\label{eq:qm}
\end{equation}
with $M_\mathrm{TAMS}$ being the mass of a star at the TAMS. 

As \citet{2008A&A...488.1017G}, we assume that a fraction $\phi$ of the total mass of a binary is lost in the \hl{merging process} and that the composition of the lost material is the same as the initial composition. The latter assumption breaks down if those parts of stars lost in the merger \hl{process} are already enriched in hydrogen-burning products, \eg, by stellar wind mass loss or rotational mixing. From head-on collisions of massive MS stars, \citet{2013MNRAS.434.3497G} find $\phi=0.3q/(1+q)^2$ with $q=M_{2}/M_{1}$ being the mass ratio and $M_1$ and $M_2$ the masses of the primary and secondary star at the time of the \hl{merging}, respectively. The mass of the merger \hl{product} is then $M=(1-\phi)(M_1+M_2)$.

Using the above definitions and assumptions, the average hydrogen mass fraction of the merger product, $\left<X\right>$, follows from $M \left<X\right> = M_1 \left<X\right>_1 + M_2 \left<X\right>_2 - \phi (M_1 + M_2) X_0$,
\begin{align}
\frac{\left<X\right>}{X_0} = {}& \frac{(1-Q_\mathrm{m,1}f_1)(1-Q_\mathrm{c,1}f_1-\phi)}{(1-\phi)\left[(1-Q_\mathrm{m,1}f_1) + (1-Q_\mathrm{m,2}f_2)q \right]} \nonumber \\
& + \frac{(1-Q_\mathrm{m,2}f_2)(1-Q_\mathrm{c,2}f_2-\phi)q}{(1-\phi)\left[(1-Q_\mathrm{m,1}f_1) + (1-Q_\mathrm{m,2}f_2)q \right]}.
\label{eq:avrg-x-merger-wind-mass-loss}
\end{align}
The indices 1 and 2 refer to the primary and secondary star, respectively. The apparent fractional MS age of the merger product, $f_\mathrm{app}$, is then related to the fractional MS age, $f_\mathrm{i}$, of a genuine single star of initial mass $M_\mathrm{i}$ that has the same mass ($M = (1 - Q_\mathrm{m,i} f_\mathrm{i}) M_\mathrm{i}$) and average hydrogen mass fraction ($\left< X \right> = (1 - Q_\mathrm{c,i} f_\mathrm{i}) X_0$) as the merger product. Introducing the parameter $\alpha$ that accounts for potential extra mixing of fresh fuel into the core of the merger product \citep[see also][]{2008A&A...488.1017G}, we find for the apparent fractional MS age
\begin{equation}
f_\mathrm{app} = \frac{f_\mathrm{i}}{\alpha} = \frac{1 - \left< X \right>/X_0}{\alpha Q_\mathrm{c,i}}.
\label{eq:fapp-wind-mass-loss}
\end{equation}

In case of negligible stellar wind mass loss, \ie $Q_\mathrm{m,1} = Q_\mathrm{m,2} = Q_\mathrm{m,i} \approx 0$, we recover the average hydrogen mass fraction and the apparent fractional MS age of the merger \hl{products} of \citet{2008A&A...488.1017G},
\begin{equation}
\frac{\left< X \right>}{X_0} = 1 - \frac{1}{1-\phi} \frac{Q_\mathrm{c,1} f_1 + Q_\mathrm{c,2} f_2 q}{1+q},
\label{eq:avrg-x-merger}
\end{equation}
\begin{equation}
f_\mathrm{app} = \frac{1}{\alpha Q_\mathrm{c}(M)} \frac{1}{1-\phi} \frac{Q_\mathrm{c,1}f_1 + Q_\mathrm{c,2}f_2 q}{1+q}.
\label{eq:fapp}
\end{equation}
\citet{2008A&A...488.1017G} and \citet{2013MNRAS.434.3497G} conduct smoothed-particle hydrodynamical (SPH) simulations of head-on collisions of massive MS stars and calibrate the extra mixing parameter $\alpha$. To that end they import the structure of their SPH merger products into a 1D stellar evolutionary code and determine the remaining MS lifetime of the merged stars, $t_\mathrm{MS}=\tau_\mathrm{MS}(1-f_\mathrm{app})$. Knowing $t_\mathrm{MS}$ from the evolution of the SPH merger \hl{products}, \citet{2008A&A...488.1017G} infer a mixing parameter of $\alpha=1.67$ for low-mass mergers (\hl{less than} $1.2\,\msun$) and \citet{2013MNRAS.434.3497G} $\alpha=1.14$ for high-mass mergers ($5\text{--}40\,\msun$) at solar metallicity. We adopt $\alpha=1.14$ for our rejuvenation prescriptions of high-mass MS mergers.

In order to understand the basic behaviour of rejuvenation, it is instructive to consider the case of equal-mass mergers with negligible wind mass loss ($M_1=M_2$, $q=1$, $f_1=f_2=f$ and $Q_\mathrm{m,1} = Q_\mathrm{m,2} = Q_\mathrm{m,i} = 0$). Equation~(\ref{eq:fapp}) then reads
\begin{equation}
\frac{f_\mathrm{app}}{f} = \frac{1}{1-\phi} \cdot \frac{Q_\mathrm{c}(M_1)}{\alpha Q_\mathrm{c}(M)}.
\label{eq:fapp-q-1}
\end{equation}
This equation shows that rejuvenation is the stronger the lower the fraction of lost mass, $\phi$, the larger the increase of the effective core mass fraction, $Q_\mathrm{c}(M)/Q_\mathrm{c}(M_1)$, and the larger the mixing into the core, $\alpha$. Equation~(\ref{eq:fapp-q-1}) further illustrates that the absolute rejuvenation, $\Delta f=f-f_\mathrm{app}\propto f$, is larger in more evolved mergers. 

\begin{figure*}
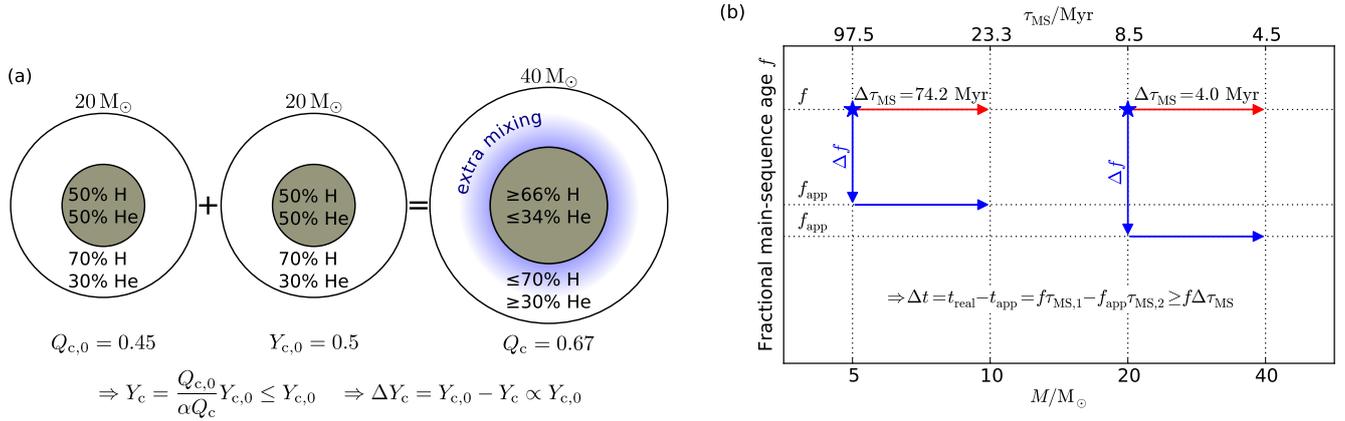

\begin{centering}
\includegraphics[width=0.49\textwidth]{{{rejuvenation-1}}}
\hspace{0.02\textwidth}
\includegraphics[width=0.47\textwidth]{{{rejuvenation-2}}}
\par\end{centering}
\caption{Cartoons illustrating the two contributions to the overall rejuvenation in the case of equal-mass mergers with no mass loss. Stars rejuvenate because of (1) the mixing of fresh fuel into the core and (2) shorter lifetimes connected with more massive stars (not to scale; see main text for details). In (a) we depict the real rejuvenation in a $10+10\,\msun$ merger, \ie the reduction of the core helium mass fraction $Y_\mathrm{c}$. Convective cores are indicated by the grey shaded regions. The parameter $\alpha$ describes the extra-mixing that occurs during the merger process. The real rejuvenation is stronger in more massive stars because they have larger effective convective cores, $Q_\mathrm{c}$, and in older stars because they have larger core helium mass fractions. In (b) we show the overall rejuvenation in the fractional-MS-age vs. stellar-mass plane for mergers of $5+5\,\msun$ and $20+20\,\msun$ for the cases of rejuvenating ($\Delta f=f-f_\mathrm{app}>0$; blue arrows) and not rejuvenating ($\Delta f=0$; red arrows) the cores of the merger products. The overall rejuvenation, $\Delta t=t_\mathrm{real} - t_\mathrm{app}$, is at least $\Delta t=f\left[\tau_\mathrm{MS}(M)- \tau_\mathrm{MS}(2M)\right]=f\Delta \tau_\mathrm{MS}$ (no real rejuvenation), \ie it is of the order of the nuclear timescale of stars. Also, the overall rejuvenation is stronger in relatively older stars (larger fractional MS ages $f$) and in lower-mass stars (larger MS-lifetime differences $\Delta \tau_\mathrm{MS}$). In terms of mass ratios $q=M_2/M_1$, rejuvenation tends towards zero for very low-mass companions.}
\label{fig:rejuvenation-cartoon}
\end{figure*}

Physically, an equal-mass merger at $f=1$, \ie a merger of two stars with pure helium cores, only rejuvenates by the additional mixing of fresh fuel into the newly formed core. However, the detailed merger simulations of \citet{2013MNRAS.434.3497G} show the formation of a thick hydrogen-burning shell around the inert helium cores that burns hydrogen for a significant fraction of time: the merger of a $20.0+19.8\,\msun$ binary at core hydrogen exhaustion of the primary produces a merger product that burns hydrogen in a thick shell for \hl{approximately} $1.4\,\mathrm{Myr}$, a duration that corresponds to about 18\% of the MS lifetime of the former primary star. This phase of thick hydrogen shell burning takes place in the blue part of the Hertzsprung--Russell (HR) diagram where genuine single stars undergo core hydrogen burning and Eqs.~(\ref{eq:fapp-wind-mass-loss}) and~(\ref{eq:fapp}) still give a good approximation to the remaining hydrogen-burning time for such merger products \citep{2013MNRAS.434.3497G}. However, the evolutionary track of such merger products in the HR diagram do no longer resemble those of single stars. The internal structure does not adjust to that of a single star with the same mass and the apparent age inferred for such merger \hl{products} from their position, \eg, in the HR diagram might be different from what we can predict from the above simplified model (see also Sec.~\ref{sec:systematic-uncertainties}). 

Let $t_\mathrm{real}=f_1\tau_\mathrm{MS}(M_1)=f_2\tau_\mathrm{MS}(M_2)$ be the real and $t_\mathrm{app}=f_\mathrm{app}\tau_\mathrm{MS}(M)$ the apparent age of the merger. The absolute amount by which the merger apparently rejuvenates, $\Delta t = t_\mathrm{real} - t_\mathrm{app}$, is then given by
\begin{equation}
\Delta t = f_1\tau_\mathrm{MS}(M_1) - f_\mathrm{app}\tau_\mathrm{MS}(M),
\label{eq:abs-dt}
\end{equation}
and the relative amount by
\begin{equation}
\frac{\Delta t}{t_\mathrm{real}} = 1 - \frac{f_\mathrm{app}}{f_1} \frac{\tau_\mathrm{MS}(M)}{\tau_\mathrm{MS}(M_1)}.
\label{eq:rel-dt}
\end{equation}
Equation~(\ref{eq:rel-dt}) demonstrates that the rejuvenation of MS mergers is of the order of the real age of the merger, \ie of the order of the nuclear timescale of stars, and further shows the two different contributions to the overall rejuvenation (see also Fig.~\ref{fig:rejuvenation-cartoon}). The first factor, $f_\mathrm{app}/f_1$, describes the real rejuvenation in terms of the mixing of fresh fuel into the core of the merger product whereas the second factor, $\tau_\mathrm{MS}(M)/\tau_\mathrm{MS}(M_1)$, describes the apparent rejuvenation because of the shorter lifetimes associated with more massive stars ($\tau_\mathrm{MS}(M)<\tau_\mathrm{MS}(M_1)$ because $M>M_1$).

\begin{table}
\caption{Initial mass $M_\mathrm{ini}$, TAMS mass $M_\mathrm{TAMS}$, effective convective core mass fraction $Q_c$ and MS lifetime $\tau_\mathrm{MS}$ of \citet{2011A&A...530A.115B} solar-metallicity stellar models used to compute the rejuvenation of merger \hl{products}.}
\label{tab:stellar-models}
\vspace{-2mm}
\begin{center}
\begin{tabular}{cccc}
\toprule
$M_\mathrm{ini}/\msun$ & $M_\mathrm{TAMS}/\msun$ & $Q_c$ & $\tau_\mathrm{MS}/\mathrm{Myr}$ \\
\midrule
\midrule
3.0 & 3.0 & 0.23 & 340.4 \\
5.0 & 5.0 & 0.26 & 97.5 \\
10.0 & 9.9 & 0.33 & 23.2 \\
20.0 & 19.3 & 0.45 & 8.5 \\
40.0 & 32.9 & 0.67 & 4.5 \\
50.0 & 33.6 & 0.86 & 3.9 \\
60.0 & 39.5 & 0.88 & 3.5 \\
80.0 & 45.2 & 0.96 & 3.1 \\
100.0 & 49.4 & 0.98 & 2.8 \\
\bottomrule
\end{tabular}
\end{center}
\end{table}

\begin{figure}
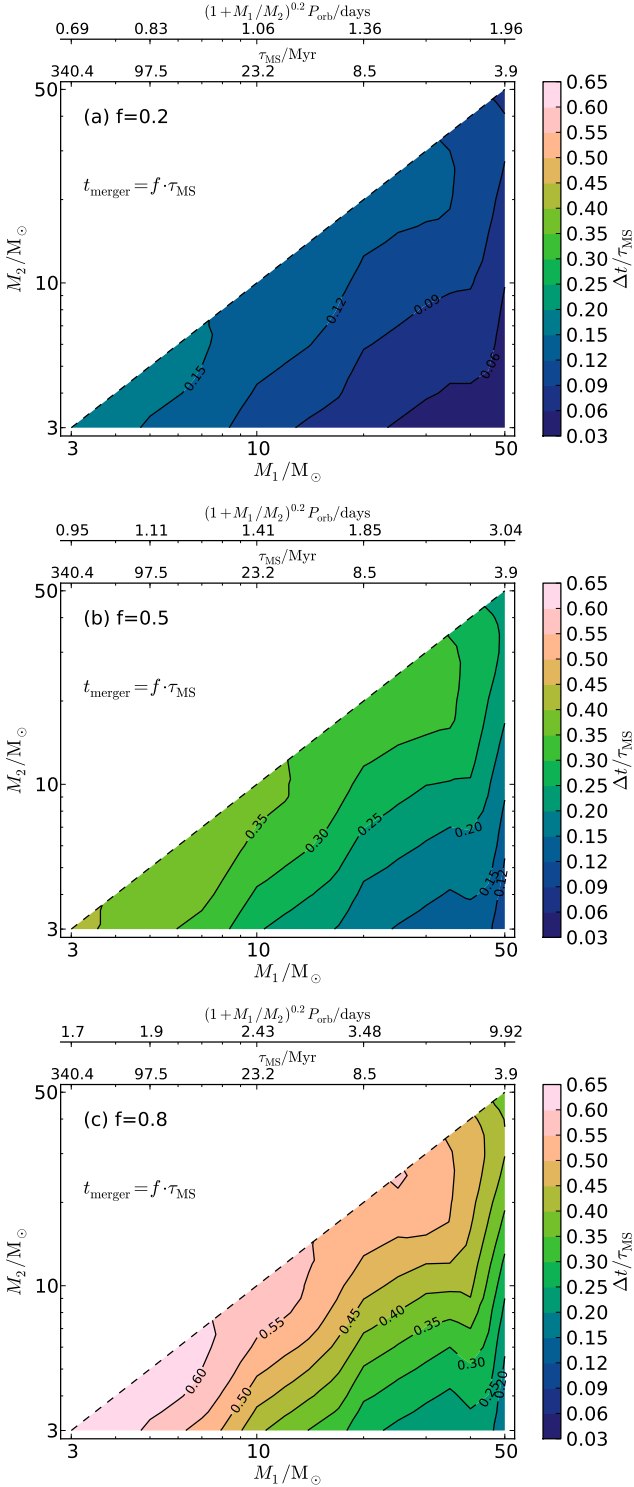

\begin{centering}
\includegraphics[width=0.48\textwidth]{{{merger-f-0.2}}}
\includegraphics[width=0.48\textwidth]{{{merger-f-0.5}}}
\includegraphics[width=0.48\textwidth]{{{merger-f-0.8}}}
\par\end{centering}
\caption{Apparent rejuvenation, $\Delta t/\tau_\mathrm{MS}$, as a function of initial primary mass, $M_\mathrm{i,1}$, and secondary mass, $M_\mathrm{i,2}$, for three different merger times: the stars merge at a fractional MS age of (a) $f=0.2$, (b) $f=0.5$ and (c) $f=0.8$ of the primary star. The upper two x-axes indicate the MS lifetime of the primary star, $\tau_\mathrm{MS}$, and the orbital period, $P_\mathrm{orb}$, of the binary at the time of the merger. The orbital periods depend slightly on the present-day mass ratio of the binary.}
\label{fig:m1-m2-dage}
\end{figure}

\begin{figure}
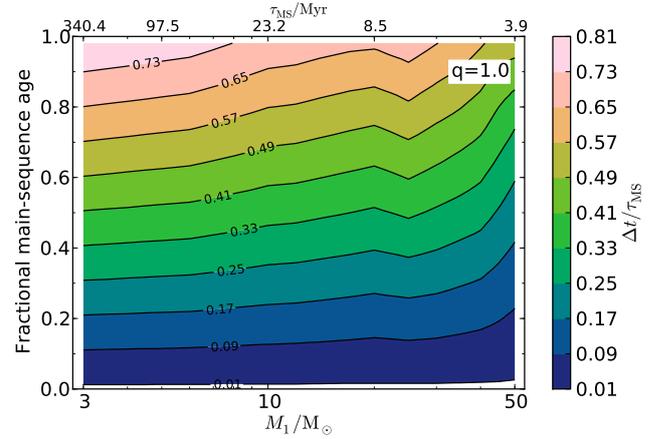

\begin{centering}
\includegraphics[width=0.48\textwidth]{{{merger-q-1.0}}}
\par\end{centering}
\caption{Apparent rejuvenation, $\Delta t/\tau_\mathrm{MS}$, as a function of the initial mass, $M_\mathrm{i,1}$, and fractional MS age of the primary star. The initial mass ratio $q_\mathrm{i}=M_\mathrm{i,2}/M_\mathrm{i,1}$ is set to 1. The top \hl{$x$-axis} gives the MS lifetime of the primary stars from which the time of the merger follows as $t_\mathrm{merger}=f\cdot \tau_\mathrm{MS}$.}
\label{fig:m1-f-dage}
\end{figure}

We now use the solar-metallicity ($X_0=0.72739$) stellar models of \citet{2011A&A...530A.115B} to compute the rejuvenation of binary mergers. In Tab.~\ref{tab:stellar-models}, we provide the initial and TAMS masses, the MS lifetimes and the effective convective core masses of these models. Within this table, we interpolate linearly in mass to obtain $M_\mathrm{TAMS}$ and $Q_c$, and logarithmically in mass to get $\tau_\mathrm{MS}$. With these data and interpolations we map the rejuvenation, $\Delta t/\tau_\mathrm{MS}$, as a function of primary and secondary mass for fractional MS ages of the primary star of $f_1 \equiv f=0.2$, $0.5$ and $0.8$ in Fig.~\ref{fig:m1-m2-dage} and as a function of primary mass and fractional MS age for a fixed mass ratio of $q=1$ in Fig.~\ref{fig:m1-f-dage}. The MS lifetimes of the primary stars, $\tau_\mathrm{MS}$, are indicated on the upper x-axes of Figs.~\ref{fig:m1-m2-dage} and~\ref{fig:m1-f-dage} from which the time of the merger follows as $t_\mathrm{merger}=f \tau_\mathrm{MS}$. At the time of the merger, the primary star fills its Roche lobe. Approximating the Roche lobe radius $R_\mathrm{L}$ by
\begin{equation}
\frac{R_\mathrm{L}}{a} = 0.44 \frac{q^{-1/3}}{(1+q^{-1})^{1/5}} \quad \text{for} \quad 0.1\leq q^{-1} \leq 10
\label{eq:RL}
\end{equation}
\citep{2006epbm.book.....E} with $a$ being the orbital separation, we find for the orbital period, $P_\mathrm{orb}$,
\begin{equation}
\left(1+\frac{M_1}{M_2}\right)^{1/5} P_\mathrm{orb} = 2\pi \sqrt{\frac{R_\mathrm{L}^3}{G M_1 0.44^3}},
\label{eq:Porb}
\end{equation}
where $G$ is the gravitational constant. Note that $M_1$ and $M_2$ are present-day masses. From further interpolations of stellar radii, we compute the orbital periods of the merging binaries following Eq.~(\ref{eq:Porb}). Because $(1+M_1/M_2)^{1/5}$ is a weak function of the mass ratio (\hl{approximately} $1.15$ for $M_1=M_2$ and $1.62$ for $M_1=10 M_2$), the orbital periods given on the second top \hl{$x$-axis} in Fig.~\ref{fig:m1-m2-dage} are close to the orbital periods of the binaries at the time of the merger.

As already discussed above and illustrated in Fig.~\ref{fig:rejuvenation-cartoon}, the overall rejuvenation is stronger in more evolved stars (cf. Fig.~\ref{fig:m1-f-dage} and Eq.~\ref{eq:fapp-q-1}). For fixed fractional MS ages, the rejuvenation is more in binaries with larger mass ratios, \ie in more massive binaries, mainly because of shorter lifetimes associated with more massive mergers. Fixing the fractional MS age and the mass ratio (\eg dashed diagonal lines for equal-mass mergers in Fig.~\ref{fig:m1-m2-dage}), the overall rejuvenation is more in less massive binaries. For example, in a $3+3\,\msun$ merger at $f=0.5$, $f_\mathrm{app}/f_1=0.82$ and $\tau_\mathrm{MS}(M)/\tau_\mathrm{MS}(M_1)=0.23$ whereas for a $30+30\,\msun$ merger $f_\mathrm{app}/f_1=0.60$ and $\tau_\mathrm{MS}(M)/\tau_\mathrm{MS}(M_1)=0.60$. The real rejuvenation, the mixing of fresh fuel into the core of the merger product, is greater in more massive stars (smaller $f_\mathrm{app}/f_1$ values) because $Q_\mathrm{c}$ increases more steeply with mass at high masses than at low masses (cf. Eq.~\ref{eq:fapp-q-1} and Tab.~\ref{tab:stellar-models}). The overall apparent rejuvenation, however, is more in lower-mass stars (smaller $\tau_\mathrm{MS}(M)/\tau_\mathrm{MS}(M_1)$ values) because of the larger mass-luminosity exponent and hence larger differences in MS lifetimes.

From Fig.~\ref{fig:m1-f-dage} we can infer the age of the apparently youngest merger product in coeval star clusters of various ages. The apparently youngest and at the same time one of the most massive merger products stems from equal-mass mergers at $f\approx 1$. In a $3.9\,\mathrm{Myr}$ star cluster, the age of the apparently youngest merger \hl{product} is about $1.9\,\mathrm{Myr}$ and in a $340\,\mathrm{Myr}$ star cluster it is about $68\,\mathrm{Myr}$. These numbers are comparable to those found by \citet{2015ApJ...805...20S} and highlight that binary products can look significantly younger than they really are, potentially biasing inferred cluster ages towards too young ages when neglecting that the most massive and hence most luminous stars are likely products of binary mass transfer \citep{2014ApJ...780..117S,2015ApJ...805...20S}.

\subsection{Model uncertainties}\label{sec:model-uncertainties}

Our rejuvenation model has essentially two physical parameters, the mixing parameter $\alpha$ and the fraction of mass lost in a merger $\phi$, and further three parameters set by the stellar models, the fraction of mass lost on the MS $Q_\mathrm{m}$, the effective core mass fraction $Q_\mathrm{c}$ and the MS lifetime $\tau_\mathrm{MS}$. We varied these parameters and report the deviations of the rejuvenation with respect to our standard model, $(\Delta t_\mathrm{standard} - \Delta t)/\Delta t_\mathrm{standard}$, in Tab.~\ref{tab:model-uncertainties}. Varying the effective core mass fraction by a factor that is independent of mass will not affect the rejuvenation in our models (see Eqs.~\ref{eq:avrg-x-merger-wind-mass-loss},~\ref{eq:fapp-wind-mass-loss} and~\ref{eq:fapp}). Any mass-independent change of the MS lifetime will linearly affect the rejuvenation for fixed fractional MS ages (see Eq.~\ref{eq:abs-dt}): prolonging MS lifetimes by, \eg, 10\%, leads to 10\% increases of the rejuvenation and vice versa. Overall the deviations are of the order of $10\text{--}20\%$ and in general larger for higher-mass binaries.

\begin{table}
\caption{Uncertainties in the amount of rejuvenation because of uncertainties in the mixing parameter $\alpha$, the fraction of mass lost in a merger $\phi$, the fraction of mass lost by stellar winds $Q_\mathrm{m}$ and switching wind mass losses totally off, $Q_\mathrm{m}=0.0$. Tabulated values are relative deviations from our standard model ($[\Delta t_\mathrm{standard} - \Delta t]/\Delta t_\mathrm{standard}$) with negative deviations indicating more rejuvenation and positive less. The quoted numbers are for $f=0.8$.}
\label{tab:model-uncertainties}
\vspace{-2mm}
\begin{center}
\begin{tabular}{ccccccc}
\toprule
$M_1=$ & \multicolumn{2}{c}{$3\,\msun$} & \multicolumn{2}{c}{$10\,\msun$} & \multicolumn{2}{c}{$50\,\msun$} \\
$q=$ & $1.0$ & $0.2$ & $1.0$ & $0.2$ & $1.0$ & $0.2$ \\
\midrule
\midrule
$\alpha+0.2$ & $-4\%$ & -- & $-6\%$ & $-21\%$ & $-17\%$ & $-31\%$ \\
$\alpha=1.0$ & $+3\%$ & -- & $+6\%$ & $+20\%$ & $+16\%$ & $+28\%$ \\
\midrule
$\phi\cdot 2$ & $+8\%$ & -- & $+12\%$ & $+20\%$ & $+12\%$ & $+17\%$ \\
$\phi/2$ & $-3\%$ & -- & $-5\%$ & $-9\%$ & $-6\%$ & $-8\%$ \\
\midrule
$Q_\mathrm{m} \cdot 1.2$ & $\pm 0\%$ & -- & $\pm 0\%$ & $\pm 0\%$ & $-12\%$ & $-2\%$ \\
$Q_\mathrm{m} \cdot 0.8$ & $\pm 0\%$ & -- & $\pm 0\%$ & $\pm 0\%$ & $+7\%$ & $+2\%$ \\
\midrule
$Q_\mathrm{m}=0$ & $\pm 0\%$ & -- & $+1\%$ & $+1\%$ & $+18\%$ & $+9\%$ \\
\bottomrule
\end{tabular}
\end{center}
\end{table}

In low-mass stars (\hl{less than about} $1.2\,\msun$) the mixing parameter $\alpha$ is $1.67$ \citep{2008A&A...488.1017G} whereas it is $1.14$ in high-mass stars \citep{2013MNRAS.434.3497G}. Increasing (decreasing) the mixing parameter results in more (less) rejuvenation. Setting $\alpha=1.0$ corresponds to no mixing of fresh fuel into the core of the merger \hl{product}. There is a mass dependence in the deviations because the mixing parameter always assumes that a fraction of the effective core mass is mixed into the core as fresh fuel. Hence the absolute amount of mixed fresh fuel is larger for more massive stars because they have larger convective cores. 

The SPH simulations of \citet{2013MNRAS.434.3497G} are head-on collisions. On the one hand, one may expect less mass loss in more gentle mergers of orbiting stars but, on the other hand, there is a lot of orbital angular momentum that may make the merged star rotate rapidly, thereby enhancing mass loss. Increasing (decreasing) the fraction of mass lost by a factor of 2 reduces (enhances) the rejuvenation. Modifying the total mass of the merger product affects the rejuvenation in two ways: (i) it influences the rejuvenation directly via changing $\phi$ in Eqs.~(\ref{eq:avrg-x-merger-wind-mass-loss}) and~(\ref{eq:fapp}) and (ii) indirectly via changes in the effective convective core sizes and the MS lifetimes. Overall the changes are such that the rejuvenation varies more in higher-mass binaries.

Stellar-wind mass-loss rates are uncertain. Increasing (decreasing) the fraction of mass lost during the MS evolution leads to more (less) rejuvenation because the initial mass of the genuine single star that matches the mass of the merger \hl{product} and the average hydrogen mass fraction is larger (smaller) for more (less) mass lost and the corresponding MS lifetimes are hence shorter (longer). Enhancing/reducing the mass loss by 20\% mainly affects the most massive binaries. Also the deviations from our standard model are smaller for younger fractional MS ages. Switching wind mass loss totally off ($Q_\mathrm{m}=0$) reduces the rejuvenation by at most 20\% for the most massive binaries considered here and leaves the rejuvenation nearly unchanged in intermediate-mass binaries (\hl{approximately} $10\,\msun$).

\subsection{Systematic uncertainties}\label{sec:systematic-uncertainties}

The age differences of our model relate to the available fuel in the cores of stars. This age difference is not the same an observer might infer from the surface properties of stars. This can cause systematic differences between the predicted rejuvenation of our models and that inferred from observations. Such differences are expected to be larger if the internal structure of \hl{merged stars} deviate significantly from that of genuine single stars of the same mass. In the models of \citet{2013MNRAS.434.3497G} this applies to mergers of stars near the end of core hydrogen burning that undergo a phase of thick hydrogen shell burning.

Generally, merger products have larger average mean molecular weights, $\mu$, and are therefore over-luminous and have larger radii (hence smaller surface gravities) compared to genuine single stars of the same mass (\eg see Fig.~7 in \citealt{2013MNRAS.434.3497G}). Thus we expect that apparent ages inferred from comparing the positions of mergers to single-star models in HR diagrams are younger than what our models predict because of shorter MS lifetimes associated with more luminous stars. In contrast, we expect older apparent ages when comparing mergers to single-star models in the so-called Kiel diagram (effective temperature vs. surface gravity diagram) because of older ages associated with stars of smaller gravities. Inferring ages by matching effective temperatures, luminosities and surface gravities simultaneously to stellar models may cancel some of the biases.

\subsection{Inferring merger progenitor properties}\label{sec:inferring-progenitor-properties}

The rejuvenation derived in this work can be used to put constraints on pre-merger binaries if the apparent age discrepancy of merger candidates with respect to comparison clocks, \eg the age of a host star cluster, is known from observations. Imagine there is an apparently single star with an inferred mass of $18.4\,\msun$ and an apparent age of $8.3\,\mathrm{Myr}$ in an otherwise coeval star cluster of $16.6\,\mathrm{Myr}$. The age of the star cluster implies that the merger must have happened at $t_\mathrm{merger}<16.6\,\mathrm{Myr}$. The mass of the merger \hl{candidate} restricts the potential mass range of pre-merger binaries and the observed apparent age discrepancy can then be used to determine the mass ratio and the time when the merger occurred. In this example, a $10+10\,\msun$ merger at an age of $t_\mathrm{merger}=11.6\,\mathrm{Myr}$ ($f_1=0.5$) leaves a merger product of $18.4\,\msun$ that looks younger by $8.3\,\mathrm{Myr}$ than it really is. From the potential binary masses and the time when the merger occurred, we can further determine the orbital period of the pre-merger binary as $P_\mathrm{orb}=1.23\,\mathrm{days}$. The pre-merger binary configuration can then be compared to binary models to check whether such a binary is indeed expected to merge. According to the models of \citet{2015ApJ...805...20S}, a binary composed of two $10\,\msun$ stars in a $1.23\,\mathrm{days}$ orbit ($a=13\,\rsun$) will indeed merge.

Generally speaking, the problem is to constrain four parameters, the primary and secondary mass, the orbital period and the time when the merger occurred, from three observables, the inferred mass and apparent and real age of the merger \hl{product}. This is obviously a degenerate situation where no unique solution exists unless in special cases where it is possible to, \eg, obtain the time of the merger from the expansion age of a shell that was ejected in the merger process. Although it is generally not possible to find a unique solution, one can use statistics and additional prior knowledge such as distribution functions of, \eg, binary periods and mass ratios to evaluate the likelihood of different merger progenitors. In some cases this may allow us to exclude significant parts of the parameter space.

\section{Magnetic merger candidates}\label{sec:merger-candidates}

\hl{The procedure of constraining pre-merger binaries relies on accurate and reliable merger models, and precise age and mass determinations of the merger candidate and the comparison clock. We now apply this new technique to the magnetic merger candidates HR~2949 (Sec.~\ref{sec:hr2948-2949}) and $\tau$~Sco (Sec.~\ref{sec:tau-sco}).}

\subsection{HR~2949}\label{sec:hr2948-2949}

HR~2949 and HR~2948 form a visual pair of B-type stars at a Hipparcos distance of $139^{+24}_{-18}\,\mathrm{pc}$ \citep{2007A&A...474..653V}, separated by $7.3\,\mathrm{arcsec}$ on the sky with radial velocities identical within their error bars \citep{2015MNRAS.449.3945S}. If the two stars were to form a binary system, their orbital separation would be larger than $2\times10^5\,\rsun$. \citet{2015MNRAS.449.3945S} argue that HR~2948 and HR~2949 may not be gravitationally bound because their relative proper motions are too large, rather suggesting a chance superposition. However they caution that their analysis is not fully conclusive because of the proper motion uncertainties. If HR~2948 and HR~2949 were to form a gravitationally bound binary, they would share the same age. But even if they are not gravitationally bound, their proximity on the sky and similar radial velocities suggest that they at least formed together in the same star-forming cloud and may therefore also share the same age. In the following, we assume that both stars are born at the same time either in a gravitationally-bound multiple system or in a star cluster/association.

The more luminous and more massive star, HR~2949, is a He-weak B3p IV star with a detected surface magnetic field of $B_\mathrm{p}=2.4^{+0.3}_{-0.2}\,\mathrm{kG}$ \citep{2015MNRAS.449.3945S}. \citet{2015MNRAS.449.3945S} further report that HR~2949 appears to be about $100\,\mathrm{Myr}$ younger than the non-magnetic companion, HR~2948. The ages in \citet{2015MNRAS.449.3945S} have been determined by comparing the positions of HR~2949 and HR~2948 in the HR diagram to stellar models of \citet{2012A&A...537A.146E}. However, the isochrones in their HR diagram \citep[Fig.~4 of][]{2015MNRAS.449.3945S} have been mislabelled (and the provided age uncertainties have been underestimated), requiring a re-determination of the stellar ages. Correctly labelling their isochrones, we find that the $1\sigma$ error bars of HR~2949 extend from the $10$ to the $30\,\mathrm{Myr}$ isochrone and those of HR~2948 from the $60$ to the $100\,\mathrm{Myr}$ isochrone. Thus HR~2949 indeed appears to be 50--90\% younger than HR~2948. 

In order to quantify the age discrepancy including robust error bars, we use the Bayesian tool \bonnsai\footnote{The \bonnsai web service is available at \href{http://www.astro.uni-bonn.de/stars/bonnsai}{http://www.astro.uni-bonn.de/stars/bonnsai}.} \citep{2014A&A...570A..66S}. We simultaneously match the observed luminosities, effective temperatures, surface gravities and projected rotational velocities derived by \citet{2015MNRAS.449.3945S} to the stellar models of \citet{2011A&A...530A.115B} to determine the masses and ages of HR~2949 and HR~2948. We assume a Salpeter initial mass function \citep{1955ApJ...121..161S} as initial mass prior, a Gaussian with mean of $100$ and FWHM of $250\,\mathrm{km}\,\mathrm{s}^{-1}$ as initial rotational velocity prior \citep[cf.][]{2008A&A...479..541H}, a uniform prior in age and that the rotation axes are randomly oriented in space. We find initial (and present-day) masses of $5.8^{+0.4}_{-0.2}$ and $4.0^{+0.4}_{-0.3}\,\msun$, and ages of $27.2^{+7.9}_{-11.1}$ and $59.8^{+29.2}_{-32.2}\,\mathrm{Myr}$ for HR~2949 and HR~2948, respectively. From the full age posterior probability distributions of the two stars, we find an age difference of $35.0^{+31.3}_{-32.6}\,\mathrm{Myr}$, which means that HR~2949 appears to be younger than HR~2948 with a confidence of 87.8\%.

\begin{figure}
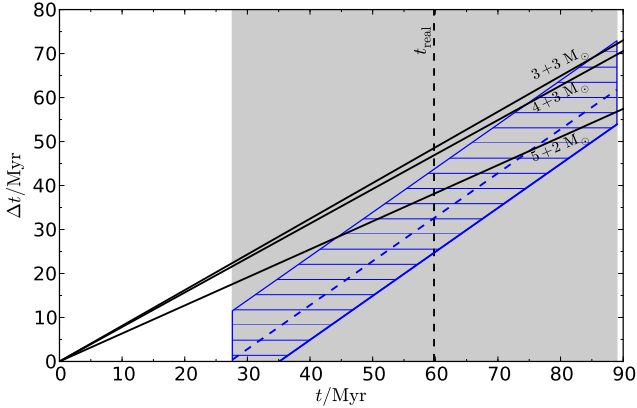

\begin{centering}
\includegraphics[width=0.48\textwidth]{{{HR2949+2948-merger-candidates}}}
\par\end{centering}
\caption{Apparent rejuvenation, $\Delta t$, as a function of the time when the merger occurred, $t$. Shown are models for $3+3\,\msun$, $4+3\,\msun$ and $5+2\,\msun$ mergers. The vertical dashed line is the age determined for HR~2948, \ie the presumed real age of the HR~2949 and HR~2948 pair, and the shaded region indicates the corresponding $1\sigma$ area. The hatched area shows the \hl{approximately} $1\sigma$ region of the inferred age discrepancy, \ie the apparent rejuvenation (note that the proper $1\sigma$ area has an elliptical shape and is not shown here for clarity).}
\label{fig:HR2949-merger-candidates}
\end{figure}

Taking this age discrepancy at face value, we investigate a potential merger origin of HR~2949 (see Sec.~\ref{sec:inferring-progenitor-properties}). In this scenario, the HR~2949 system was a triple-star system in which the inner binary merged. The inferred mass of HD~2949 requires the merger to leave a remnant of about $5.8\,\msun$. Mergers of $3+3\,\msun$, $4+3\,\msun$ and $5+2\,\msun$ stars may produce the observed mass of HR~2949 and we plot the predicted rejuvenation of such merger \hl{products} as a function of the time of the merger in Fig.~\ref{fig:HR2949-merger-candidates}. Within our models, the apparent rejuvenation is a linear function of the (fractional MS) age when the two stars merge (see Eq.~\ref{eq:fapp} and recall that $f_2=f_1 \tau_\mathrm{MS}(M_1)/\tau_\mathrm{MS}(M_2)$). We indicate the $1\sigma$ range of the real age of the merger, the age of the comparison clock HR~2948, by the shaded region in Fig.~\ref{fig:HR2949-merger-candidates} and highlight the inferred age difference of HR~2949 with respect to the comparison clock by the hatched area\footnote{We note that the outermost corners of the hatched area are not within $1\sigma$ of the observed ages of HR~2949 and HR~2948 and that a proper $1\sigma$ contour has an elliptical shape. For clarity we do not show the proper $1\sigma$ contour here.}. The inferred age difference depends on the real age and varies within the $1\sigma$ uncertainty of the age of the merger candidate HR~2949. Our merger models are able to explain the observables if they predict an age difference that is larger than the lower bound of the inferred age differences, \ie if they cross or are above the lower bound at any time younger than the real age of the HR~2949 system (\hl{younger than} $89\,\mathrm{Myr}$). In the present case all considered merger models provide the right merger remnant mass and apparent age difference when merging at ages younger than $89\,\mathrm{Myr}$.

Predicting the right amount of rejuvenation and post-merger mass within a certain time window is a necessary but not sufficient condition to explain HR~2949 by a MS merger induced by binary evolution. Depending on the initial mass ratio and orbital period, Roche-lobe overflow (RLOF) initiated during the MS evolution of the more massive donor star may lead to stable mass transfer and not a merger. The oldest possible merger time to explain the observed rejuvenation in case of the considered $3+3$, $4+3$ and $5+2\,\msun$ binaries is $89\,\mathrm{Myr}$. This translates into maximum orbital periods of $0.64$, $0.91$ and $2.20\,\mathrm{days}$, respectively, to explain all observations with such binaries (cf. Sec.~\ref{sec:mergers} and Eq.~\ref{eq:Porb}). According to the binary models of \citet{2015ApJ...805...20S} all considered binaries indeed merge. The $5+2\,\msun$ binaries merge because of the small mass ratio of $0.4$ and the $3+3$ and $4+3\,\msun$ binaries because both stars are going to overfill their Roche lobes as a result of stellar expansion on a nuclear timescale, leading to a contact phase and a subsequent merger. We therefore conclude that the properties of HR~2949 are consistent with being a merger product.

\subsection{$\tau$~Sco}\label{sec:tau-sco}

$\tau$~Sco is a bright, magnetic B0.2~V star, adjacent in the night sky to the red supergiant Antares in the zodiacal constellation of Scorpius. It is a proper motion member of the Upper Scorpius association for which \hl{\citet*{2012ApJ...746..154P}} estimate an age of $11\,\pm1\mathrm{(stat.)}\pm2\mathrm{(sys)}\,\mathrm{Myr}$. \citet{2006MNRAS.370..629D} find a surface magnetic field strength of $\tau$~Sco of about $0.5\,\mathrm{kG}$ and a surprising complex magnetic field configuration for hot stars. Because $\tau$~Sco was considered a spectral standard for B0~V stars \citep{1973ARA&A..11...29M}, many authors determined atmospheric and sometimes also fundamental parameters of $\tau$~Sco \citep[\eg][]{1991A&A...244..419K,2005A&A...441..711M,2006A&A...448..351S,2008A&A...490..793H,2012ApJ...746..154P,2014A&A...566A...7N}. \citet{2014A&A...566A...7N} note that the inferred apparent age of $\tau$~Sco does not coincide with that of the Upper Sco association and suggest that $\tau$~Sco may be a merger \hl{product}.

Analogously to Sec.~\ref{sec:hr2948-2949}, we use the Bayesian code \bonnsai \citep{2014A&A...570A..66S} to determine the mass and age of $\tau$~Sco including robust error bars by matching observed effective temperatures, surface gravities, luminosities and projected rotational velocities simultaneously to the stellar models of \citet{2011A&A...530A.115B}. The choice of priors is the same as in Sec.~\ref{sec:hr2948-2949}. We use three sets of observables, those of \citet[][M05]{2005A&A...441..711M}, \citet[][SD06]{2006A&A...448..351S} and \citet[][NP14]{2014A&A...566A...7N}, and summarize the stellar parameters and determined masses, ages, age differences to the age of Upper~Sco and the probability that $\tau$~Sco is younger than Upper~Sco in Tab.~\ref{tab:tau-sco}. The apparent age of $\tau$~Sco is in all cases significantly younger than that of the Upper~Sco association.

\begin{table}
\caption{Stellar parameters of $\tau$~Sco. Effective temperatures $T_\mathrm{eff}$, surface gravites $\log g$, luminosities $\log L/\lsun$ and projected equatorial rotational velocities $v\sin i$ are from \citet[][M05]{2005A&A...441..711M}, \citet[][SD06]{2006A&A...448..351S} and \citet[][NP14]{2014A&A...566A...7N}. The ages and masses are derived using \bonnsai applying the same stellar models and priors as for the HR~2949 system in Sec.~\ref{sec:hr2948-2949}. The inferred initial and present-day masses, $M$, are the same. The age difference to that of Upper~Sco, $\Delta t$, and the probability $p$ that the age difference is greater than zero are computed using the full posterior distributions of the inferred stellar ages \hl{and taking the uncertainty in the inferred age of the Upper~Sco association into account. The probability $p$ is therefore a measure of our confidence that the age of $\tau$~Sco is indeed younger than that of Upper~Sco and also accounts for a certain finite duration of star formation in Upper~Sco through the age uncertainty of Upper~Sco inferred by \citet{2012ApJ...746..154P}.} All error bars are $1\sigma$ uncertainties.}
\label{tab:tau-sco}
\vspace{-2mm}
\begin{center}
\begin{tabular}{cccc}
\toprule
 & M05 & SD06 & NP14 \\
\midrule
\midrule
$T_\mathrm{eff}/\mathrm{K}$ & $31900^{+500}_{-800}$ & $32000\pm1000$ & $32000\pm300$ \\
$\log g/\mathrm{cm}\,\mathrm{s}^{-2}$ & $4.15^{+0.09}_{-0.14}$ & $4.0\pm0.1$ & $4.30\pm0.05$ \\
$\log L/\lsun$ & $4.39\pm0.09$ & $4.47\pm0.13$ & $4.33\pm0.06$ \\
$v\sin i/\mathrm{km}\,\mathrm{s}^{-1}$ & $5$ (adopted) & $\leq 13$ & $4\pm1$ \\
\midrule
$M/\msun$ & $16.0^{+0.8}_{-0.6}$ & $16.8^{+1.0}_{-1.1}$ & $16.0^{+0.4}_{-0.4}$ \\
$\mathrm{Age}/\mathrm{Myr}$ & $2.0^{+1.3}_{-1.3}$ & $4.1^{+1.1}_{-1.5}$ & $0.0^{+0.6}_{-0.0}$ \\
\midrule
$\Delta t/\mathrm{Myr}$ & $8.8^{+2.6}_{-2.5}$ & $7.2^{+2.6}_{-2.6}$ & $10.6^{+2.2}_{-2.3}$ \\
$p (\Delta t > 0)$ & $>99.9\%$ & $99.8\%$ & $>99.9\%$ \\
\bottomrule
\end{tabular}
\end{center}
\end{table}

As in Sec.~\ref{sec:hr2948-2949}, we check whether the found age discrepancy is consistent with the rejuvenation predicted in our merger models (Fig.~\ref{fig:tau-Sco-merger-candidates}). To match the observed mass of $\tau$~Sco, we consider mergers of $8.5+8.5$, $10+7$, $11+6$, $12+5$, $13+4$ and $14+3\,\msun$ binaries\footnote{The merger of a $9+8\,\msun$ binary is not shown because the predicted rejuvenation is nearly the same as that of the equal-mass $8.5+8.5\,\msun$ mergers.}. As in Fig.~\ref{fig:HR2949-merger-candidates}, the shaded region indicates the age range of the comparison clock and hence the real age of $\tau$~Sco (the age of the Upper~Sco association) and the hatched areas the approximate $1\sigma$ regions of the age differences for the three sets of stellar parameters M05, SD06 and NP14. 

\begin{figure}
\begin{centering}
\includegraphics[width=0.48\textwidth]{{{tau-Sco-merger-candidates}}}
\par\end{centering}
\caption{As Fig.~\ref{fig:HR2949-merger-candidates} but for $\tau$~Sco. The merger models are for $8.5+8.5$, $10+7\,\msun$, $11+6$, $12+5$, $13+4$ and $14+3\,\msun$ binaries. The age of the Upper Sco association is given by the vertical dashed line \citep{2012ApJ...746..154P}. The hatched areas denote the approximate $1\sigma$ areas of the apparent age differences of $\tau$~Sco with respect to the ages derived from the stellar parameters of \citet[][M05]{2005A&A...441..711M}, \citet[][SD06]{2006A&A...448..351S} and \citet[][NP14]{2014A&A...566A...7N}.}
\label{fig:tau-Sco-merger-candidates}
\end{figure}

\hl{All models can explain the observables given the stellar parameters of SD06, \ie the models} predict a rejuvenation larger than the lower bound of the hatched area. Strictly speaking, the $13+4\,\msun$ mergers predict a rejuvenation that is not compatible with the observations within the $1\sigma$ uncertainties. We therefore conclude that the minimum mass ratio required to explain the SD06 observations is about $0.3$.

In the case of the stellar parameters of NP14, none of our models is able to explain the apparent age discrepancy. An increase of at least 25\% of the rejuvenation is needed to explain the observed age discrepancy with the $8.5+8.5\,\msun$ merger. Given that the uncertainties of our predictions for the rejuvenation of mergers of $10\,\msun$ stars are of the order of $5\text{--}10\%$ (Tab.~\ref{tab:model-uncertainties}), it seems difficult to explain the apparent young age of $\tau$~Sco inferred from the stellar parameters of NP14 with our models. 

The set of stellar parameters of M05 suggests an age difference that can be explained by the merger models of, \eg, $8.5+8.5$, $10+7$ and $11+6\,\msun$ binaries. Mergers of binaries of the same total mass but smaller mass ratios predict too little rejuvenation. From Fig.~\ref{fig:tau-Sco-merger-candidates} we find that, in order to explain the observed age difference, the $8.5+8.5\,\msun$ merger must occur at an age between $7.5$ and $12.5\,\mathrm{Myr}$, which translates into an orbital period range of $0.89\text{--}1.03\,\mathrm{days}$. The age intervals are approximately $7.8\text{--}10.8$ and $8.2\text{--}9.5\,\mathrm{Myr}$ and the corresponding orbital period ranges $1.00\text{--}1.14$ and $1.07\text{--}1.24\,\mathrm{days}$ for the $10+7$ and $11+6\,\msun$ merger models, respectively. Given these orbital periods and primary and secondary masses, all binaries indeed merge during the MS evolution according to the models of \citet{2015ApJ...805...20S} because of contact phases caused by the nuclear expansion of the secondary stars. A consequence of the potential age ranges of the mergers is that all mergers would be quite recent, \ie they would have had to happen less than $1\text{--}2\,\mathrm{Myr}$ ago. \citet{2014A&A...566A...7N} note that the spin-down timescale of $\tau$~Sco because of magnetic braking exceeds their inferred young age (\hl{younger than} $2\,\mathrm{Myr}$), posing a problem for our understanding of the slow rotation of this star. Also our merger models face the same issue, suggesting more efficient spin-down in the past.

In conclusion, it is plausible to explain the age differences inferred from the stellar parameters of M05 and SD06 with our merger models of close binaries (orbital periods of about $1\,\mathrm{day}$) and mass ratios larger than $0.3\text{--}0.5$. At the same time it seems difficult to explain the large age difference derived from the stellar parameters of NP14.

\section{Discussion}\label{sec:discussion}

Unambiguous and a larger number of candidates are required to fully establish the link between strong magnetic fields and mergers. As shown in this paper, looking for age discrepancies is a promising way to identify rejuvenated binary products among magnetic stars. The best candidates are those magnetic stars for which there is a robust comparison clock. The comparison clocks can be coeval star clusters and binary and higher-order multiple systems. Promising targets are the SB2 binary V1046~Ori (HD~37017) located in the NGC~1977 star cluster and $\theta^1$~Ori~C in the Trapezium cluster (having one binary companion and other cluster members as comparison clocks) and the triple star $\zeta$~Ori (HD~37742) having two gravitationally bound comparison clocks. Further observations and careful modelling are required to determine the ages of these stars and their comparison clocks with high confidence.

\subsection{Thermal timescale processes}\label{sec:accretion}

Stellar mergers and common-envelope phases are dynamical and violent phenomena, and the differential rotation they induce is thought to be the key ingredient to generate strong magnetic fields \citep{2008MNRAS.387..897T,2009MNRAS.400L..71F,2014MNRAS.437..675W}. Also mass transfer by Roche-lobe overflow (RLOF) leads to differential rotation but seems not to be related to strong magnetic fields. Some Be stars are expected to be formed from RLOF in binaries and may be ejected into the field by the supernova explosion of the former donor star \citep[\eg][]{1991A&A...241..419P,2006csxs.book..623T}. None of the observed Be stars is found to be magnetic \citep[\eg][]{2014IAUS..302..265W}. Another example is $\theta$~Car, a single-lined spectroscopic binary with an orbital period of about $2.2\,\mathrm{days}$ in the open cluster IC~2602. The visible component is a nitrogen-enriched B star rotating with a projected rotational velocity of about $110\,\mathrm{km}\,\mathrm{s}^{-1}$ that likely accreted mass during a past RLOF phase giving rise to its blue straggler appearance \citep{2008A&A...488..287H}. No magnetic field has been detected in this post-RLOF system \citep{1979ApJ...228..809B,2008A&A...488..287H}. Also mass accretion during star formation may act on a similar timescale to that of RLOF but clearly not all stars have strong surface magnetic fields. Roche-lobe overflow and mass accretion during star formation proceed at most on a thermal timescale and it therefore seems that processes acting on near-dynamical timescales such as mergers and common-envelope phases are required to generate long-lived, strong surface magnetic fields \citep{2014IAUS..302....1L}.

A star potentially contradicting this is Plaskett's star, HD~47129. Plaskett's star is currently thought to be a massive ($M_\mathrm{tot}\sin i=92.7\pm2.7\,\msun$) O star binary with an orbital period of $14.4\,\mathrm{days}$ and a mass ratio of $M_2/M_1=1.05\pm0.05$ \citep{2008A&A...489..713L}. The rapidly rotating secondary star is thought to be magnetic \citep{2013MNRAS.428.1686G} and it has been suggested that it gained its fast rotation in a past-RLOF phase (\hl{\citealt*{1992ApJ...385..708B}}; \citealt{2008A&A...489..713L,2013MNRAS.428.1686G}). However, Plaskett's star is currently being re-investigated and its inferred properties are expected to change significantly (J.~Grunhut, private communication). For the time being, we are therefore left with a puzzling situation that requires further attention to reveal the true nature of this interesting binary.

\subsection{Observational consequences of the merger hypothesis for the origin of magnetic stars}\label{sec:obs-consequences}

The merger model for the origin of strong, large-scale magnetic fields in massive stars makes clear and testable predictions beyond the rejuvenation discussed in the previous sections. Requiring mergers to produce magnetic stars implies that there should be no magnetic star in close binaries. To be more precise, there may be rare channels to form short-period binaries with one or two magnetic stars (\eg from capturing magnetic stars during the star formation process or later in cluster like environments) but the incidence of magnetic stars in close binaries is predicted to be significantly lower than in apparently single stars or wide binaries. Wide binaries such as the potential binary system HR~2949 discussed in Sec.~\ref{sec:hr2948-2949} could have been born as triple stars where the inner binary merged. Searches for magnetic OBA stars in binaries indeed show that there is a significant dearth of magnetic stars in close binaries \citep{2002A&A...394..151C,2015IAUS..307..330A,2015arXiv150200226N}, supporting the merger hypothesis.

Binary population synthesis simulations predict that the rate of MS mergers increases with mass \citep[\eg][]{2014ApJ...782....7D,2015ApJ...805...20S}. In their standard model, \citet{2014ApJ...782....7D} find a merger fraction of 8\% in a population of stars more luminous than $10^4\,\lsun$ (\hl{roughly corresponding to} OB stars) and 12\% in stars more luminous than $10^5\,\lsun$ (O stars). However, given that the present-day population of magnetic massive stars would be a mixture of pre-MS and MS mergers, the overall trend of the incidence of magnetic MS stars with mass is not readily obvious. It may be expected that the incidence of magnetic stars is greater in MS than in pre-MS stars if magnetic fields observed in the pre-MS phase prevail into the MS and if they do not disappear from the surface of stars, \eg by decaying.

The fraction of MS merger \hl{products} is expected to be highest in a population of massive blue straggler stars. In the models of \citet{2015ApJ...805...20S} MS mergers make up about 30\% of the blue straggler star population in young (\hl{less than about} $100\,\mathrm{Myr}$), coeval stellar populations (this does not include blue stragglers formed by dynamical interactions/collisions in dense clusters). So, if MS mergers indeed contribute to the magnetic massive star population, we expect a higher magnetic field incidence in massive blue straggler stars.

Surface nitrogen enrichment may be achieved by various mechanisms, for example by rotationally and magnetically induced mixing, but it is also a prediction of the suggested merger channel. \citet{2013MNRAS.434.3497G} find that the surfaces of merger \hl{products} of binaries more massive than about $5\text{--}10\,\msun$ show significant nitrogen enrichment. Furthermore they find that the surface nitrogen enrichment is stronger in more massive and more evolved binary merger progenitors. Pre-MS stars are not hot and dense enough in their cores to activate the CN(O) cycle and pre-MS mergers therefore cannot have nitrogen-enriched surfaces. Hence, if mixing processes other than merger mixing are negligible, the ratio of nitrogen-enriched to nitrogen-normal magnetic stars may be used to gain insights into the fraction of MS mergers among magnetic stars. In that regard it is interesting to note that \citet{2006A&A...457..651M} and \hl{\citet*{2008A&A...481..453M}} find a correlation between slow rotation, surface nitrogen enrichment and magnetic stars (but see also \citet{2014ApJ...781...88A} who find that magnetic fields have no predictive power for surface nitrogen enrichment in Galactic massive stars). Magnetic fields and/or stellar mergers may therefore help to explain the slowly rotating, nitrogen rich stars in the Hunter diagram of LMC B stars that are defying the predictions of state-of-the-art, rotating single-star models \citep{2008ApJ...676L..29H,2011A&A...530A.116B,2012ARA&A..50..107L}. First steps into this direction have been taken by \hl{\citet*{2011A&A...525L..11M}} and \hl{\citet*{2012MNRAS.424.2358P}}, who suggest that magnetic braking and the associated mixing may contribute to the group of slowly rotating, nitrogen-enriched stars.

Mergers are expected to eject mass because of the huge surplus of angular momentum. \citet{2013MNRAS.434.3497G} find that, depending on the mass ratio of the \hl{merging binary}, about $2\text{--}9\%$ of the total mass is ejected. In mergers of massive stars this may produce ejecta of a few solar masses, implying that some magnetic stars should be surrounded by massive nebulae\footnote{Ejected nebulae disperse and are therefore only visible for a limited amount of time (probably \hl{less than about} $10^5\,\mathrm{yr}$).}. As suggested by \citet{2012ARA&A..50..107L}, the nitrogen-enriched (5 times solar), young ($3000\,\mathrm{yr}$), massive ($2\,\msun$), expanding ($350\,\mathrm{km}\,\mathrm{s}^{-1}$), bipolar nebula RCW~107 surrounding the magnetic O6.5f?p star HD~148937 \citep{1987A&A...175..208L} may well be the ejecta of a merger. Also the B[e] supergiant in the wide binary system R4 ($P_\mathrm{orb}\approx 21\,\mathrm{yr}$) in the Small Magellanic Cloud is suggested to be a merger \hl{product} and is surrounded by a young ($1.2\times 10^4\,\mathrm{yr}$), nitrogen-enriched, bipolar nebula expanding with a velocity of about $100\,\mathrm{km}\,\mathrm{s}^{-1}$ \citep{2000AJ....119.1352P}. However the large distance to this star makes it hard to establish whether the star is magnetic. Furthermore some red luminous novae, \eg V4332~Sgr, V838~Mon and V1309~Sco, may be stellar mergers \citep{1999AJ....118.1034M,2002A&A...389L..51M,2011A&A...528A.114T}. In particular the eruption of V838~Mon may have involved rather massive B stars (\eg \citealt{2005A&A...434.1107M}; \hl{\citealt*{2005A&A...441.1099T}}), making this target a promising candidate for probing the idea of magnetic-field generation in massive-star mergers.

\section{Conclusions}\label{sec:conclusions}

Merging MS stars leads to rejuvenation because of mixing of fresh fuel into the cores of merger products and shorter lifetimes associated with more massive stars. We find that the rejuvenation is of the order of the nuclear timescale of stars, implying that merger products can look substantially younger than they really are. Using the results of SPH merger models of \citet{2013MNRAS.434.3497G}, we show that the rejuvenation is the stronger the more evolved the progenitors, the lower the masses of binaries and the larger the mass ratios. Given our models, it is possible to identify MS merger \hl{products} by their rejuvenation and to put constraints on the merger progenitor from the mass, the apparent age and the real age of the merger product. The latter can be inferred from comparison clocks such as a binary companions and other coeval cluster members. 

Merging of MS and pre-MS stars has been suggested to lead to the strong, large-scale magnetic fields found in about 10\% of MS OBA and pre-MS Herbig Ae/Be stars \citep{2009MNRAS.400L..71F,2012ARA&A..50..107L,2014IAUS..302....1L,2014MNRAS.437..675W}. If this hypothesis is true, the magnetic stars originating from MS mergers should look significantly younger than other coeval comparison stars. We find clear age discrepancies in HR~2949 with respect to its potential binary companion HR~2948 and in $\tau$~Sco with respect to the Upper~Sco association. The inferred age discrepancies of both magnetic stars, their masses and real ages are consistent with our merger models, suggesting that these stars may indeed be merger \hl{products} that obtained their magnetic fields in the merging process. 

However, because of uncertainties regarding the coevality of HR~2948 and HR~2949, and published stellar parameters for $\tau$~Sco, further rejuvenated magnetic stars need to be identified to substantiate the hypothesis of merging for the formation of magnetic stars. Finding MS merger candidates by their young ages relies on robust comparison clocks and we highlight a few magnetic stars for future investigations for which good comparison clocks are available. Searching for apparent age discrepancies in magnetic stars is therefore a promising way to investigate the origin of strong, large-scale magnetic fields in OBA stars and to understand the evolution and final fates of this intriguing class of stars.

\section*{Acknowledgements}

FRNS thanks the Hintze Family Charitable Foundation and Christ Church College for his Hintze and postdoctoral research fellowships, respectively. \hl{The authors thank the anonymous referee for helpful comments.}




\bibliographystyle{mnras}

\bsp	
\label{lastpage}
\end{document}